\documentclass{KapProc}
\usepackage{psfig}
\normallatexbib
\begin{document}
\articletitle{Homodyne characterization \\ of active optical media}
\author{G. Mauro D'Ariano, Matteo G. A. Paris and Massimiliano F. Sacchi}
\affil{Theoretical Quantum Optics Group, INFM and Dipartimento di Fisica ``A. Volta'' \\ 
Universit\`a di Pavia, via Bassi 6, I-27100 Pavia, ITALY} %\email{paris@unipv}
\begin{abstract}
An effective maximum likelihood method is suggested to characterize 
the absorption/amplification properties of active optical media 
through homodyne detection. 
\end{abstract}
The quantum characterization of optical media is an important issue 
 in modern optical technology, since the noise in optical 
communications and measurements is ultimately of quantum origin.  
For negligible saturation effects, the propagation of an optical
signal in an active media is governed by the master equation
\begin{eqnarray}
\dot \varrho =G_1\, L[a]\,\varrho +G_2\, L[a^\dag ]\,\varrho
\;,\label{meq}
\end{eqnarray}
where $\varrho$ is the density matrix describing the quantum state 
of the signal mode  $a$ and $L[O]$ denotes the Lindblad superoperator 
$L[O] A=O\,A\,O^\dag -\frac 12 O^\dag O\,A - \frac 12 \,A\,O^\dag O$. 
If we model the propagation as the interaction of a traveling wave 
single-mode $a$  with a system of N identical two-level atoms, 
then the absorption $G_1=\gamma N_1$ and amplification $G_2=\gamma N_2$
parameters are related to the number $N_1$ and $N_2$ of atoms in the 
lower and upper level respectively. The quantity $\gamma$ is a rate of the
order of the atomic linewidth \cite{mandel}, and the propagation gain (or
deamplification) is given by ${\cal G}=\exp [(G_2-G_1)\: t]$. \par 
An active medium described by the master equation (\ref{meq}) represents a 
kind of phase-insensitive optical device. In this paper, we want 
to evaluate the parameters $G_1$ and $G_2$ by the maximum-likelihood (ML) 
estimation applied to data coming from random phase 
homodyne detection on the signal exiting the medium.  
The present investigation is motivated by the fact that ML approach 
has been already successfully applied to estimation of 
the whole quantum state \cite{mxl} as well as to determination of 
some parameters of interest in quantum optics \cite{pxl}. 
\par
Let us start by reviewing the ML approach. Let $p(x | \lambda)$ 
the probability density of a random variable $x$, 
conditioned to the value of the
parameter $\lambda$.  The analytical form of $p$ is known, but the true value of 
the parameter $\lambda$ is unknown, and should be estimated from the
result of a measurement of $x$.  In our case $\lambda$ is the couple 
of parameters $G_1$ and $G_2$, and $p$ is the probability density 
of (random phase) homodyne data. Let $x_1, x_2, ..., x_N$ be a random
sample of size $N$. The joint probability density of the independent
random variable $x_1, x_2, ..., x_N$ (the global probability of the
sample) is given by
\begin{eqnarray}
{\cal L}(\lambda)= \Pi_{k=1}^N \: p(x_k |\lambda)
\label{likdef}\;,
\end{eqnarray}
and is called the likelihood function of the given data sample. 
The ma\-xi\-mum-like\-lihood estimator  of the parameter 
$\lambda$ is defined as the quantity $\lambda_{ml}$ 
that maximizes ${\cal L} (\lambda)$ for variations of $\lambda$.
Since the likelihood is positive this is equivalent to maximize 
\begin{eqnarray}
L(\lambda) = \log {\cal L} (\lambda) = \sum_{k=1}^N \log p(x_k | \lambda)
\label{loglikfun}\;
\end{eqnarray}
which is the so-called log-likelihood function. 
\par
Using the Wigner representation of Eq. (\ref{meq}) one can easily solve the
corresponding Fokker-Plank equation for the Wigner function $W(\alpha
,\alpha ^*;t)$. One obtains \cite{ours}
\begin{eqnarray}
W(\alpha,\alpha ^*;t)=
\int {\mbox{d}^2\beta \over \pi \delta ^2}\exp\left( -{|\alpha -g\beta |^2
\over \delta ^2}\right) W(\beta,\beta ^*;0)
\;\label{conv}
\end{eqnarray}
with $g=e^{-Qt}$, $2Q=(G_1-G_2)$, and $\delta ^2=(G_1+G_2)(1-g^2)/(4Q)$.
The theoretical homodyne probability at phase $\phi $ is simply
obtained as the following marginal distribution
\begin{eqnarray}
p(x;\phi)= \int\mbox{d}\hbox{Im}\alpha \, W(\alpha e^{i\phi},\alpha ^*
e^{-i\phi})
\qquad x=\hbox{Re}(\alpha )\;.
\end{eqnarray}
\par For input coherent state with amplitude $\alpha 
_0$, one has $W(\beta,\beta ^*;0)=\frac 2\pi \exp (-2|\beta -\alpha
_0|^2)$ and the convolution in Eq. (\ref{conv}) gives 
\begin{eqnarray}
W(\alpha,\alpha ^*;t)=\frac {1}{\pi (\delta ^2+g^2/2)}\exp
\left( -{|\alpha -g\alpha _0 |^2
\over \delta ^2 +g^2/2}\right) 
\;.\label{wout}
\end{eqnarray}
The corresponding theoretical homodyne distribution is then given by
\begin{eqnarray}
p(x;\phi )=\frac {1}{\sqrt{\pi(\delta ^2 +g^2/2)}}\exp\left\{-\frac 
{1}{\delta ^2+g^2/2}\left[ x-g\hbox{Re}(\alpha _0\,e^{-i\phi })\right]^2 
\right\}\;.\label{x1}
\end{eqnarray}
For non-unit quantum efficiency $\eta < 1$, one has the replacement 
\begin{eqnarray}
\delta ^2+\frac {g^2}{2} 
\longrightarrow \delta ^2+\frac{g^2}{2} + \frac{1-\eta }{2\eta}\;.
\end{eqnarray}
We applied the ML approach to determine $G_1$ and $G_2$ starting
from random phase homodyne detection [$\phi$ in Eq. (\ref{x1}) randomly 
distributed in $(0,\pi)$].
As a input reference signal we used coherent state of fixed known
amplitude. Notice that the use of coherent states is not simply a
matter of computational and experimental convenience. In fact, there
is no advantage in using e.g. squeezed states, because of the
phase-insensitive character of the device. Compare, on the contrary,
the case of phase estimation in Ref. \cite{pxl}.  
Some results from Monte Carlo simulated experiments for both the
absorption  ($G_1 > G_2$) and 
the amplification ($G_1 < G_2$) regime are shown in Table
\ref{t:uno}. Notice also that the case $G_1=G_2$ corresponds to the
estimation of Gaussian noise, 
since one has the solution of Eq. (\ref{meq}) in the form 
\begin{eqnarray}
\varrho (t)= \int \frac{\mbox{d}^2\beta}{\pi G_1 t}\,e^{-\frac{|\beta |^2}{G_1 
t}}\,D(\beta )\varrho D^\dag (\beta ) \;,
\end{eqnarray}
where $D(\beta )=e^{\beta a^\dag -\beta ^* a}$ denotes the
displacement operator.
\begin{table}[h]\begin{center}
\begin{tabular}{|c|c|c|c|c|c|} \hline
$G_1$ & $G_2$  & $(G_1)_{ML}$& $(\delta G_1)_{ML}$& $(G_2)_{ML}$& 
$(\delta G_2)_{ML}$ \\ \hline
3. & 1.&  2.97250576 & 0.03489146 & 0.96966708 & 0.03299910 \\ \hline
  3. & 2.&  2.93669546 & 0.04629955 & 1.94330199 & 0.04412476\\ \hline
  3. & 3.&  3.03023643 & 0.07376747 & 3.03199992 & 0.07122468\\ \hline
  3. & 4.&  2.98543015 & 0.09926873 & 3.98150430 & 0.09763157\\ \hline
  3. & 5.&  3.16888784 & 0.06556872 & 5.15783291 & 0.06240839\\ \hline
\end{tabular}
\end{center}
\caption{Maximum likelihood estimation of loss ($G_1$) and gain 
($G_2$) parameters of master equation (\ref{meq}). 
Input coherent state with amplitude $\alpha _0 =4$ has been used, 
along with $N=10^4 $ homodyne data with quantum efficiency $\eta 
= 0.6$, collected after an effective time $t=1$. $\delta G_1$ 
and $\delta G_2$ represent the statistical error of the
estimation.\label{t:uno}}
\end{table} \par
\begin{figure}[ht]
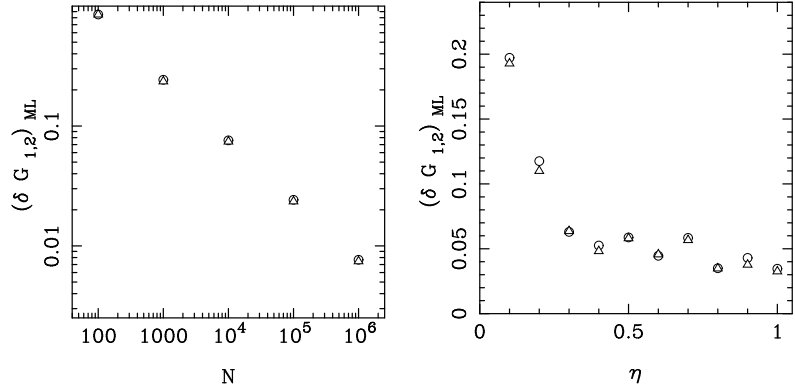

\begin{tabular}{cc}
\psfig{file=errsudatalog.ps,width=5cm}&
\psfig{file=errsueta.ps,width=5cm}  
\end{tabular}
\caption{Behaviour of the statistical error $(\delta G_1)_{ML}$
(circle) and $(\delta G_2)_{ML}$ (triangle) versus number of homodyne
data $N$ (left, notice the scaling $\delta G \propto N^{-0.509}$) 
and quantum efficiency of homodyne detector (right) in the 
maximum likelihood estimation of $G_1$ and $G_2$ in the
master equation (\ref{meq}). Parameters (left): $G_1=3$,  
$G_2=5$, $\eta =0.6$, $\alpha _0 =5$, $t=1$.  
Parameters (right): number of data $N= 5\times 10^3$, $G_1=2$,  
$G_2=1$, $\alpha_0 =8$, $t=1$.}\label{f:err}
\end{figure}

In Fig. \ref{f:err} we show the behavior of the 
statistical errors on the maxlik determination of the parameters as a 
function of the number of homodyne data and the quantum efficiency of 
photodetectors. 
The robustness of the method to low quantum efficiency $\eta $ is a
feature of the maximum-likelihood technique \cite{mxl,pxl}. In the
present case, however, it is not 
surprising [see Fig. (\ref{f:err})], because quantum efficiency
itself can be described by master equation (\ref{meq}) \cite{ours}. 
Notice the inverse square root behaviour of the statistical errors
versus the number $N$ of data in the sample, according to the central  
limit theorem.\par
In conclusion, we applied the maximum-likelihood estimation approach to the 
characterization of linear active optical media through homodyne detection.
The resulting method is efficient and provides a precise determination of the 
absorption and amplification parameters of the master equation using small 
homodyne data sample.
\par\acknowledgment This work has been supported by the Italian Ministero 
dell'Universit\`a e della Ricerca Scientifica e Tecnologica (MURST)
 under the co-sponsored project 1999 {\em Quantum Information
Transmission And Processing: Quantum Teleportation And Error Correction}. 

\end{document}